\documentclass[aps,prl,twocolumn,superscriptaddress,grouapedaddress]{revtex4}
\usepackage{graphicx}
\usepackage{latexsym}
\usepackage{amsmath}
\usepackage{amsfonts}
\usepackage{amssymb}
\usepackage{color}
\usepackage{units}
\newcommand{\be}{\begin{equation}}
\newcommand{\ee}{\end{equation}}
\newcommand{\bea}{\begin{eqnarray}}
\newcommand{\eea}{\end{eqnarray}}
\usepackage[parfill]{parskip}
\usepackage{orcidlink}
\begin{document}
\title{Shape, temperature and density interplay in depletion forces} 
\author{Itay Azizi \orcidlink{0000-0003-2939-4421}} 
\affiliation{Donostia International Physics Center, 20018 Donostia, Spain} %
\affiliation{Department of Physics and Institute of Nanotechnology and Advanced Materials, Bar-Ilan University, Ramat-Gan 52900, Israel} %
\address{Correspondence: itay.azizi@gmail.com} %
\begin{abstract}
Via numerical simulations and analytical calculations, depletion forces are studied in mixtures of small and big particles that interact via soft repulsive potentials. While big particles are spherical, small particles are nonspherical with shapes that vary gradually, from squares to rods via intermediate shapes. The mixtures 
are studied for a wide range of densities and temperature. Depletion forces and their resulting potentials depend on the interplay of shape, temperature and density, an argument that is elaborated qualitatively and quantitatively. While in some thermodynamic conditions, depletion potentials of distinct shapes are distinguishable, in different conditions, they are very similar. Finally, I propose novel computational models and experiments for further investigation of the effect of morphology on phase separation in and out of thermal equilibrium.
\end{abstract} 
\maketitle 
\section{Introduction}
Primary examples for mixtures of small and big particles that exhibit entropy-driven demixing transition are for spherical particles that interact via hard-core potential \cite{Dijkstra1999}. Some works on depletion studied mixtures of particles with nonspherical shapes, such as rods, lens and dumbbells 
\cite{Mao1995,Yaman1998,lang2007,sun2012,Eisenriegler2005}. Y-shaped particles can be antibody immunoglobulin-G (IgG) that consists of four peptide chains: two identical heavy chains and two identical light chains \cite{Werner1972}, or trinaphthylene that creates a NOR logic gate on a golden surface \cite{Sow2011}. Elongated particles can be bacteria or viruses, specifically, tobacco mosaic virus and fibrils of amyloid $\beta$-protein, the molecular agent at the origin of the Alzheimer disease \cite{Dogic2006,Mao1995}. \newline
\textcolor{white}{facil} Interest in mixtures of small and big particles interacting via different potentials stems from the appearance of entropy-driven demixing transition where big particles tend to form a pure phase of solid (or liquid) coexisting with a pure phase of small particles fluid. The basic mechanism behind the big particles clustering is that it frees sufficient available volume for the small particles, that such big particles confinement would increase the total entropy of the system and a phase separated state would be favored over a mixed state \cite{AO1954,AO1958}. Such demixing requires high enough size assymetry and sufficiently large densities of small and big particles \cite{Dijkstra1999}. 
\newline \textcolor{white}{facil} In my previous studies, I showed that depletion interactions can modify the transition temperature for partial freezing of big particles \cite{Azizi2020} and lead to unusual reentrant transitions \cite{Azizi2022}. Demixing depends primarily on the numbers of small and big particles that should  very large to sustain a stable phase separation. To economize the number of particles in the system, one can coarse grain the depletion interaction and study smaller systems of only big particles that interact via an effective potential that consists a two-body depletion potential. Such method was used for mixtures of small and big spheres, in three dimensions, that interact via soft-repulsive potential \cite{Cinacchi2007}. In this work, I study the forces acting on big disks by small depletants that can have one of the five following forms: disk, square, Y, hockey or rod. Therefore, studying the depletion effect as a function of particle shape, at different small particles densities and temperatures, is the  subject  of  the  present work which, for simplicity, is done in two dimensions. 
\section{Methods}
\begin{figure}[ht] 
\includegraphics[width=0.8\linewidth]{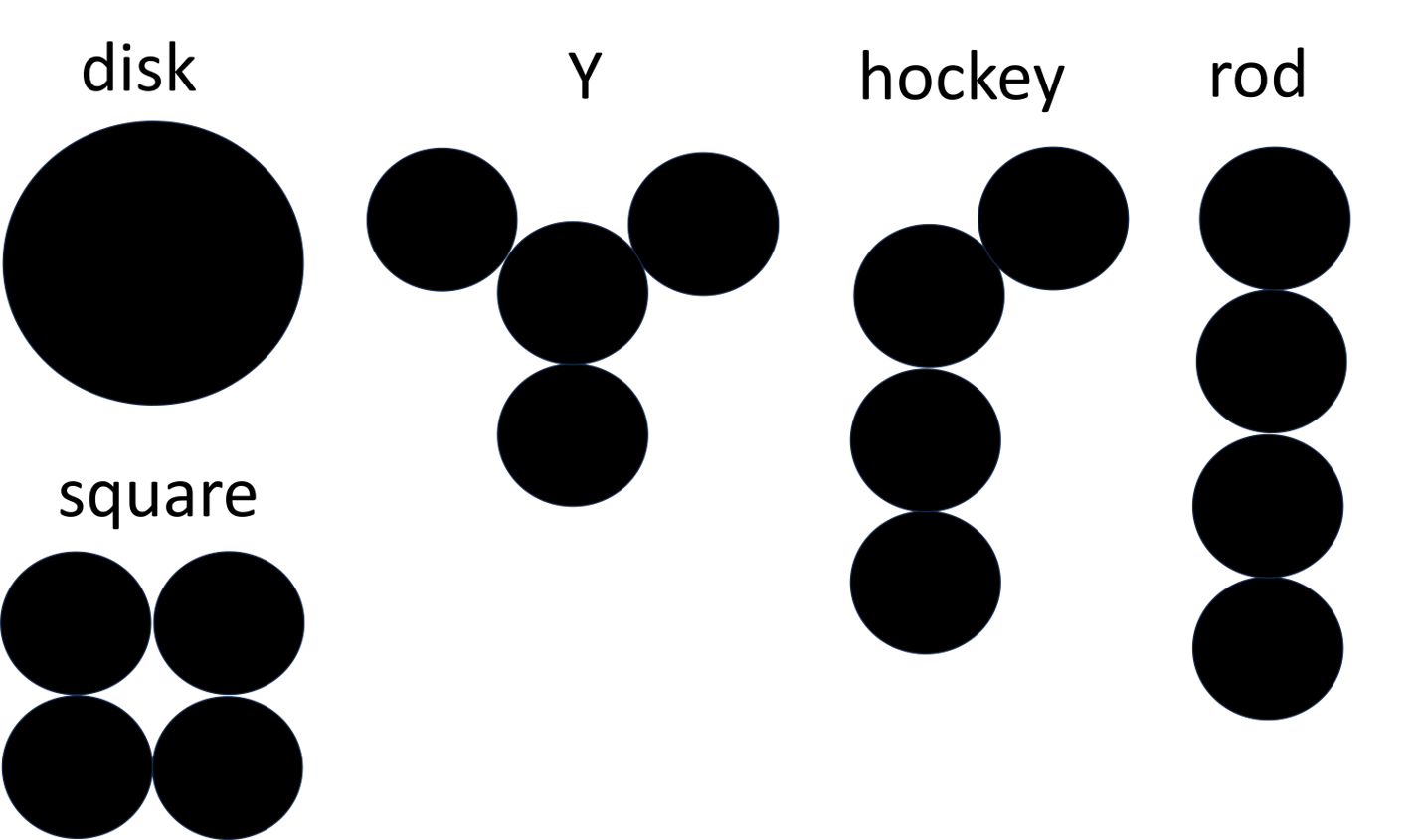}
	\caption{\label{fig:types} The depletants fluid particles can be in one of these shapes.}
\end{figure}

\begin{figure}[ht] 
\includegraphics[width=0.8\linewidth]{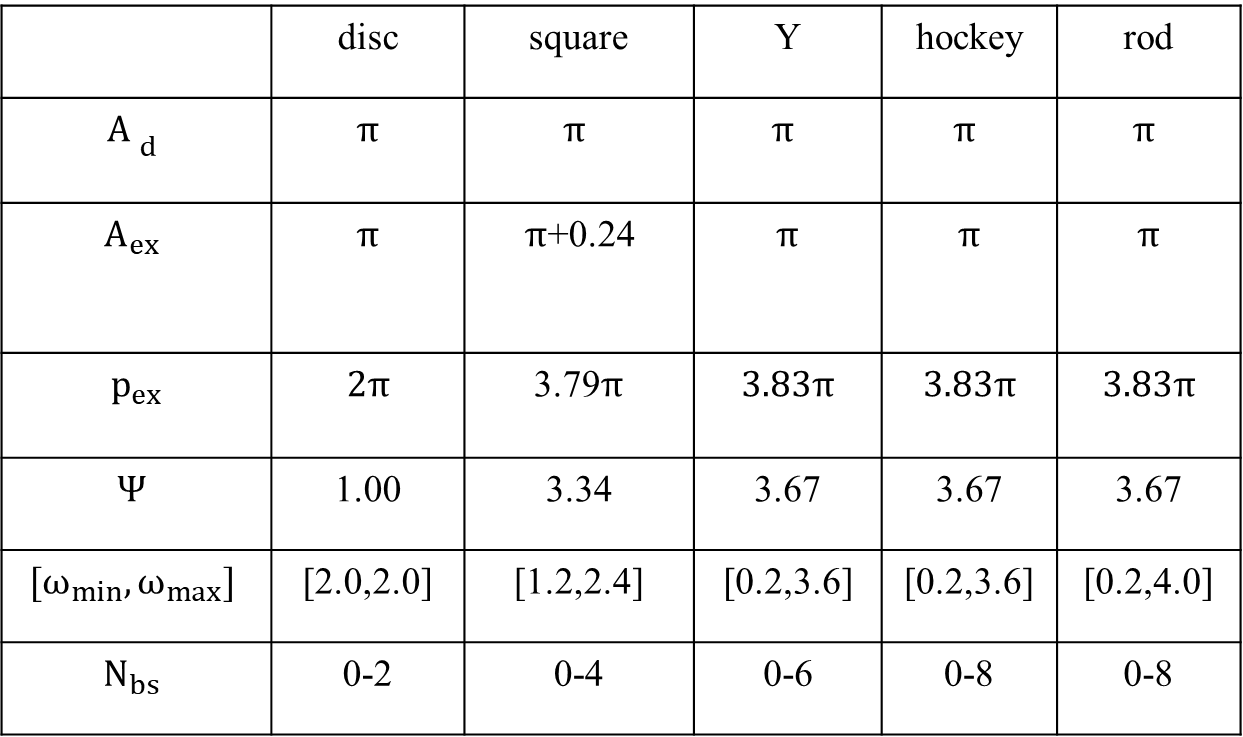}
	\caption{\label{fig:table}  Geometric properties of the particles, in length units of a small monomer diameter.}
\end{figure}

To elucidate the shape-dependent behavior of a binary mixture of big and small particles, I carried out two-dimensional Monte Carlo simulations in the canonical NVT ensemble of a pair of big particles in a fluid of small particles. 
\newline
\underline{Modelling the depletants:}
\newline
Each nonspherical depletant (square, Y, hockey or rod) is a rigid polymer that consists of four disks that are permanently attached such that the interparticle distances between the small disks are fixed, as well as the angles (see Fig.\ref{fig:types}). The building blocks are disks with a LJ diameter of $\sigma_{ss}=1$ that sets the unit of length in the system.  For instance, Y particles consist of four small disks with three equally-spaced arms by $2\pi/3$ angles and R particles are linear chains of four disks. The polymer-type depletants are comparable with disks with LJ diameter of $\sigma_{dd}=2\sigma_{ss}=2$, such that all shapes areas are equal to $4\pi(\sigma_{ss}/2)^2=\pi$. The gyration radii of different shapes are slightly different, approximately $R_g=\sigma_{ss}=1$. The big disks have a diameter of $\sigma_{bb}=10$ that sets the size ratio of small to big particles to be $q=R_g/R_{big}\approx0.2$. 
\newline
To discern between the geometric properties of the different shapes, see Fig.\ref{fig:table}. The properties are: 
\begin{itemize}
\item “Area of disks” $A_{d}$ -  the area oocupied by the building blocks.
\item “Excluded area” $A_{ex}$ – area of envelope that encircles the particle which is a sum over the partial circular contributions. 
\item “Excluded perimeter” $p_{ex}$ – perimeter of the excluded area.
\item “Roundness” $\psi$ - using a known definition:
\begin{eqnarray}
    \psi=\frac{p_{ex}^2}{4 \pi A_{ex}}
\end{eqnarray}
This ratio will be 1 for a circle and greater than 1 for non-circular shapes.
\item “Linear range” $[a_{min},a_{max}]$ – the range of minimal to maximal linear dimensions of the envelope.
\item “Interactivity” $N_{bs}$ – the number of possible interactions between small disks to big disks.
\end{itemize} 
Minimal and maximal linear dimensions provide insights relating shape to depletion strength. The maximal dimension sets the distance above which depletion vanishes, at $R=\sigma_{bb}+\omega_{max}$. This distance shrinks as temperature increases. In addition, it indicates the "interactivity", so longer polymers can repel the big disks stronger than short ones. The minimal dimension does not play important role in determination of the depletion, however, it is relevant for one-component organization of a pure phase of a single shape and this will be discussed in a follow up project that is still under investigation. The quantitative deviation from roundness agrees with visual inspection : disks are $100$ percents round, squares deviate much and Ys, hockeys and rods deviate even more due to their branched structure. 
\underline{Modelling the interactions:} 
\newline
Spherical particles in the system interact via soft repulsive potential, specifically via WCA (Weeks-Chandler-Andersen) potential, that reads 
\begin{eqnarray}
U_{ij}(r)=4\epsilon_{ij}[(\sigma_{ij}/r)^{12}-(\sigma_{ij}/r)^{6}]+\epsilon_{ij}
\label{eq:WCA_pot},
\end{eqnarray} 
with cut-off distance of $\sigma'_{ij}=2^{1/6} \sigma_{ij}\approx 1.122 \sigma_{ij}$, where $i$, $j$ are either $b$, $s$ or $d$. The sum rule is assumed for interactions between different disks, $\sigma_{ij}=(\sigma_{ii}+\sigma_{jj})/2$. Therefore, the repulsive interactions between the depletants is between their building blocks, and building blocks that belong to same depletants do not interact. Unit of energy is $\epsilon=1$ and all interaction energies are chosen to be the same $\epsilon_{ij}=\epsilon=1$ for all possible combinations of $i$ and $j$.
\underline{Temperature range:} \newline
In fluids with continuous monotonic soft repulsive pair potential, such as WCA potential, the typical interparticle distance between a pair of particles that repel each other is a decreasing function of increasing temperature. This behavior was shown at different works and particularly in two dimensions in my article \cite{Azizi2022} where I measured the radial distribution functions of WCA fluids simulated via molecular dynamics. At very low temperature, about T=0.01, this distance is maximal which is the cutoff of the WCA potential $1.122\sigma_{ij}$. At intermediate temperature, for example T=1, this distance is near $\sigma_{ij}$, and at very high temperature, about T=10, this distance is significantly below $\sigma_{ij}$ and there is visible interpenetration. Based on this temperature scale, I chose low temperature as $T=0.025$, such as it is not too low and would result in super slow equilibration. The intermediate temperature as $T=1$ and the high temperature as $T=10$. \newline
\underline{Simulation box and algorithm:} 
\newline
\begin{figure*}[htpb] 
\centering
\includegraphics[width=0.9\textwidth,height=4cm]{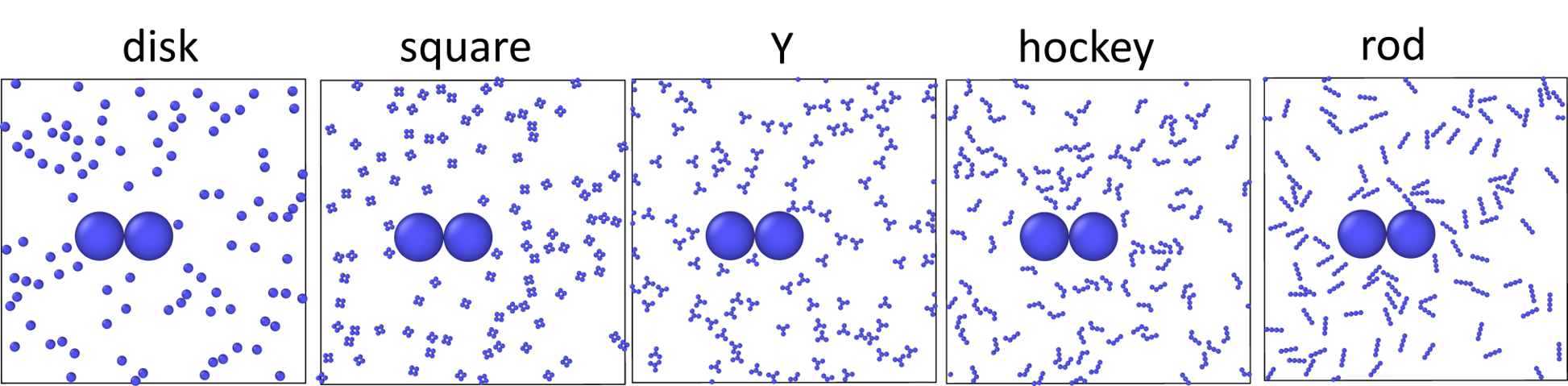}
 \caption{\label{fig:snapshots} Snapshots of steady state configurations of a pair of big disks in a fluid of depletants at different shapes that interact via WCA potential at $T=1$. At all systems shown the big disks are fixed at distance $r=10\sigma_{ss}$ and the depletant density is $\rho_{dep}=0.025\sigma_{ss}^{-2}$.}
\end{figure*}
The simulation prescription is as follows. Simulation box is a square of length $L=62 \sigma_{ss}$ with periodic boundary conditions in both $x$ and $y$ directions that contains a pair of big disks that are fixed at distance $r$ and the depletants change their positions and orientations. See Fig.\ref{fig:snapshots} for the snapshots of the simulation for the parameters: $\rho_{dep}=0.025$, $T=1$ and $r=10$. Depletant density is 
\begin{eqnarray}
\rho_{dep}=N_{dep}/L^2
\label{eq:density},
\end{eqnarray}
where $N_{dep}$ is the number of depletants. The depletant area fraction is according to our previous work \cite{Azizi2022} based on the cut-off diameter as 
\begin{eqnarray}
\phi_{dep}=4N_{dep}\pi(\frac{\sigma'_{ss}}{2})^2/L^2=4\pi(\frac{\sigma'_{ss}}{2})^2\rho_{dep}=2^{1/3}\pi\rho_{dep}
\label{eq:LJareafraction},
\end{eqnarray}
\newline
\textcolor{white}{facil}In each simulation, there is a single shape of depletants at a certain density and certain temperature. Each MC trial is an attempt to modify the system towards an equilibrium state with respect to density and temperature, that is done via the Metropolis algorithm. Repeteadly, a single depletant is randomly chosen and randomly shifted translationally and orientationally with amplitudes $\delta_t$ and $\delta_r$, respectively. The trial results in a new configuration with a possible potential energy gap $\Delta E=E_{new}-E_{old}$ which determines the approval or rejection of this move.  If $\Delta E<0$, then the move is approved; otherwise, a random number $p$ in the range $[0,1]$ is called and compared to $e^{-\Delta E/T}$. If $e^{-\Delta E/T}>p$, then this move is approved; if both conditions fail, then the move is rejected. If the move is rejected, then the system stays in the same state as before. Both the translational and rotational amplitudes were taken such that they produce an acceptance rate of about $0.3$ which is good enough to sample different configurations using my computational resources. In cases that it was more difficult to relax the system lower acceptance rates were taken.  
\newline
\textcolor{white}{facil}I was running each simulation for a large number of trials till it reached equilibrium behavior such that additional trials do not result in further evolution. The equilibrium state was tested by three criteria: (1) Relaxation to plateau values of the total number of big-small interactions ($n_{bs}$) and the the total potential energy associated with the big-small interactions ($U_{bs}$), (2) Relaxation of the radial distribution functions associated with the small-small interactions and the big-small interactions. (3) Relaxation of the average angular orientation to the value of $\pi$ signifying an average random orientation. See Fig.\ref{fig:snapshots} for the equilibrated configurations at $\rho_{dep}=0.025$, $T=1$ and $r=10$. 
\newline 
\underline{Depletion force measurement:} 
\newline 
The depletion force acting on a big particle is measured by the average over the total forces exerted by small particles in the direction of the axis connecting the two big particles:  
\begin{eqnarray}
F_{dep}(r)=\frac{1}{2}\langle \textbf{r} \cdot \sum_{j=1}^{N_s} (\textbf{F}_j^{(a)}-\textbf{F}_j^{(b)}) \rangle
\label{eq:Depletion_force},
\end{eqnarray} 
where $\textbf{r}$ is a unit vector in the direction of the axis connecting the two big particles, $\textbf{F}_j^{(a)}$ ($\textbf{F}_j^{(b)}$) is the force acting by the jth small disk on the left (right) big disk. Summation is over $N_s$ small disks that interact with the big particles. The factor of $1/2$ is added since the measurement is for a pair of big particles and the depletion force is acting on a single big particle. 
\newline
\textcolor{white}{facil}The depletion force, in the interparticle range of $[5,15]$, was measured such that each mean force is an average over 5000 configurations. At distance of approximately the sum of the diameters of a big particle and a small polymer, the mean force is zero or oscillating around zero. In such a Monte Carlo (MC) algorithm, the mean is over equilibrium configurations that are very sensitive to small changes on the surfaces of the big particles. To improve the accuracy of these measurements, I produced multiple simulations for the same thermodynamic variables. Each such simulation uses a slightly different MC step so it results in different particle configurations and is regarded as statistically independent.  
\newline
\underline{Integration of force into potential:}
\newline
For each point in my parameter space, there are specific values for particle shape, density and temperature and mean depeltion force profile as a function of distance between the big disks. This profile was fitted to a polynom of order $M$. The analytical expression of this polynom was integrated into a polynom of order $M+1$ that is the depletion potential up to a force-dependent constant. This function was shifted by a constant that ensures that above distance of $\sigma_{bb}+\omega_{max}$ the function vanishes or oscillates around zero. This means that for each point in the parameter space there was different force-dependent constant. In cases that the force profiles were not smooth, I fitted into two polynomial functions, each of different order, to allow a better fit of the force and therefore more accurate integration into a potential. 
\underline{Asakura-Oosawa (AO) depletion potential:} 
\newline
To adapt the AO approximation to depletants with new shapes, an integration was introduced over all the four particles consisting the polymers and also over another angle, namely: the orientation of the depletant. The AO potential is an analytical expression that can serve as a first order in density approximation to the depletion potential. 
\begin{figure}[ht] 
\includegraphics[width=1.0\linewidth]{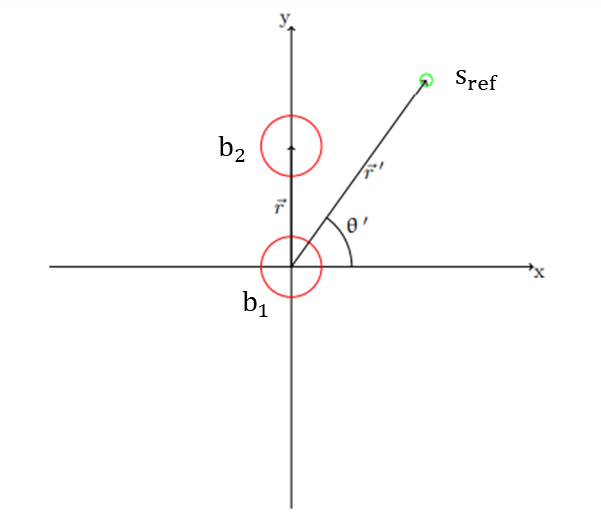}
	\caption{\label{fig:AO_illu} Illustration for the calculation of the AO potential, the reference particle is colored in green.}
\end{figure}
For dilute fluids the AO potential should be similar to the depletion potential produced by the simulations. The AO potential for depletion of two big disks by small disks reads  
\begin{eqnarray}
V_{dep}(r)=-\beta^{-1} \rho_{dep}\int_{0}^{\sigma'_{bs}}dr'r'\int_{0}^{2\pi}d\theta'f_{bs}(\textbf{r}')f_{bs}(\textbf{r}-\textbf{r}') 
\label{eq:AO_potential},
\end{eqnarray} 
where $\beta=(k_BT)^{-1}$ and $\rho_{dep}$ is the depletant density. See Fig.\ref{fig:AO_illu} for an illustration that includes the details of the vectors in this calculation. For a pair of big particles $b_1$ and $b_2$ and a small reference particle $s_{ref}$, $\textbf{r}$ is the vector from $b_1$ to $b_2$, $\textbf{r'}$ is the vector from $b_1$ to $s_{ref}$ (illustrated in green color) and $\textbf{r}'-\textbf{r}$ is the vector from 
$b_2$ to $s_{ref}$. $\theta'$ is the angle of vector $\textbf{r}'$. $f_{ij}$ is the Mayer function: $f_{ij}=1-exp(-\beta U_{ij})$. For example, for disk-shaped depletant, the AO potential vanishes at $\sigma_{bd}'=2^{1/6}(\sigma_{bb}+\omega_{max}) \approx 13.46 \sigma_{ss}$. 
Note that the AO approximation using the Mayer functions is not based on the assumption of hard-core interactions, therefore the softness of the WCA potential enters directly into this first order approximation and there is no need to use effective temperature-dependent diameters for adaptation of a hard-core potential to soft-core potential.
\section{Results}
\begin{figure}[ht] 
\includegraphics[width=1.0\linewidth]{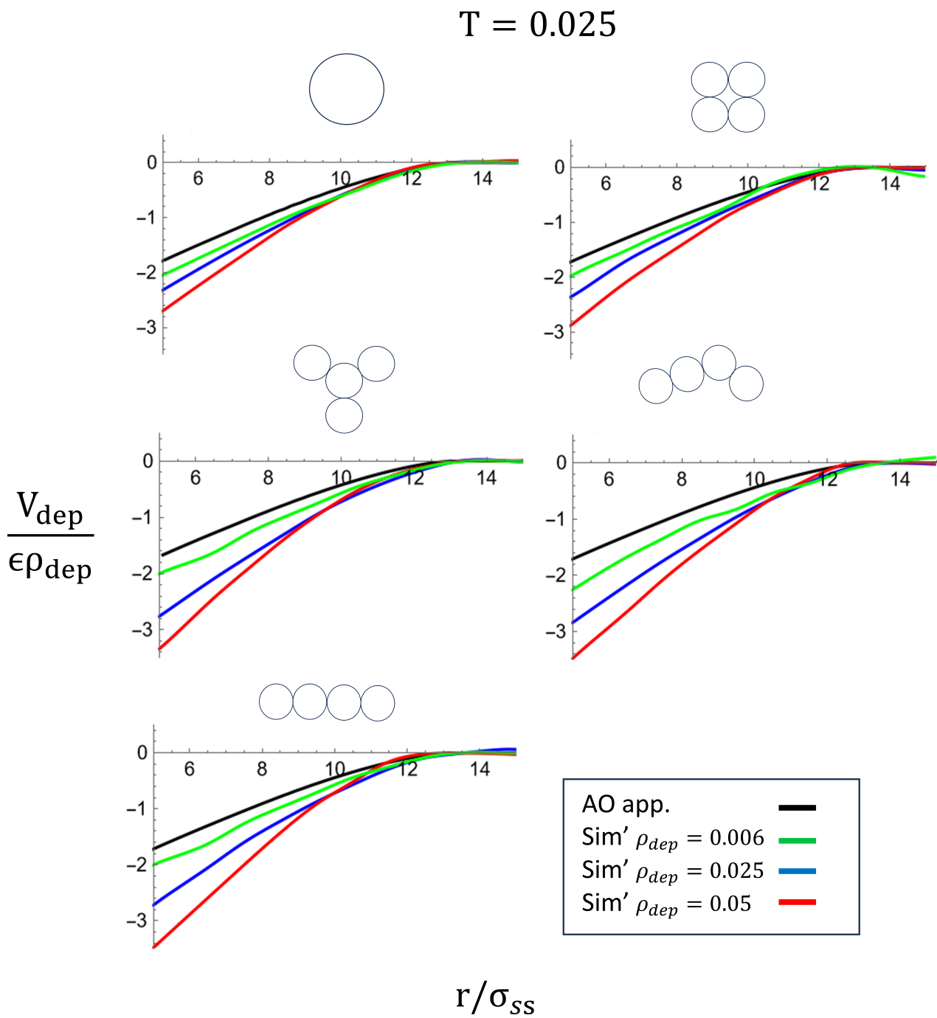}
	\caption{\label{fig:AO_SIM} Depletion potentials at temperature $T=0.025$ for an increasing depletants density as a function of particle shape and in comparison to the AO term (Eq.4).}
\end{figure}

\begin{figure}[ht] 
\includegraphics[width=1.0\linewidth]{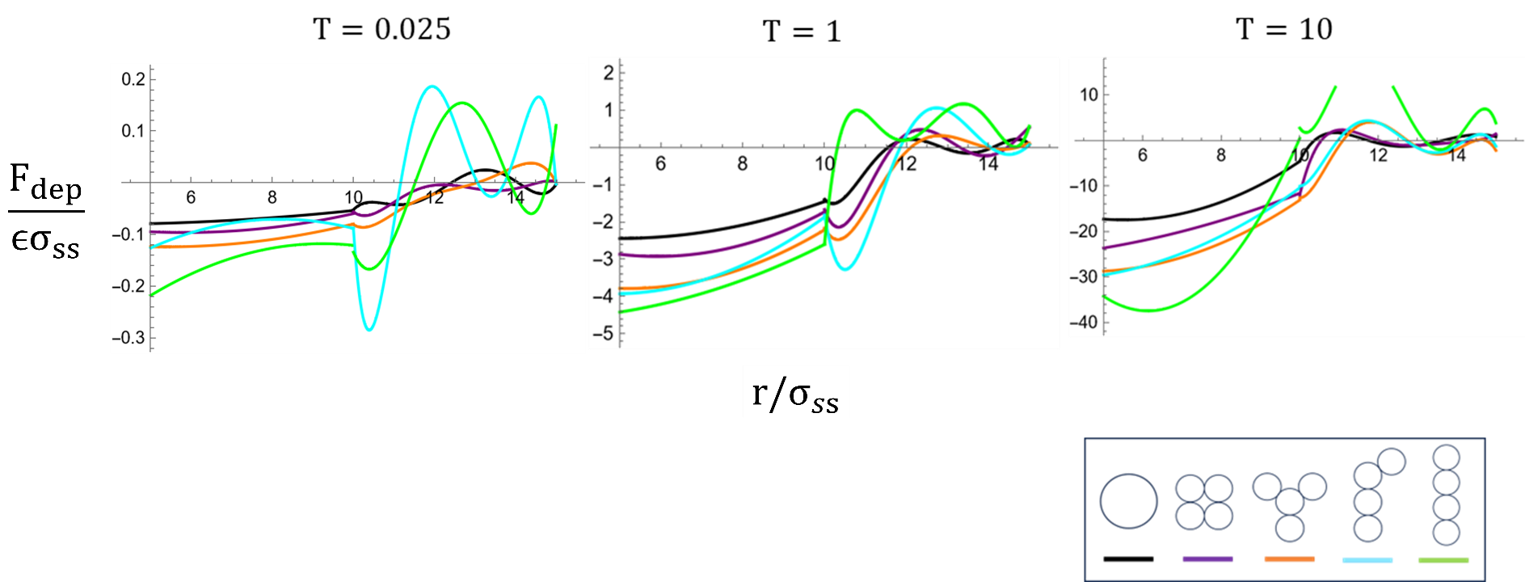}
	\caption{\label{fig:FOR2} Depletion forces with oscillating tails at high density of $\rho_{dep}=0.1$ in a wide temperature range ($T=0.025, 1, 10$) for different shapes according to the colormap.}
\end{figure}

The results are presented in two parts. The first part includes comparison between the  depletion potentials produced by simulations to their respective AO potentials. The second part includes a comparison between the depletion potentials of different shapes. In Fig.\ref{fig:AO_SIM}, the results at the hard-core limit show that for all shapes, as density increases, there is a growing deviation of the simulations from the AO term. Since the AO term is the first term in a density expansion of the potential, this result fits the theory-based prediction.
\begin{figure}[ht] 
\includegraphics[width=1.0\linewidth]{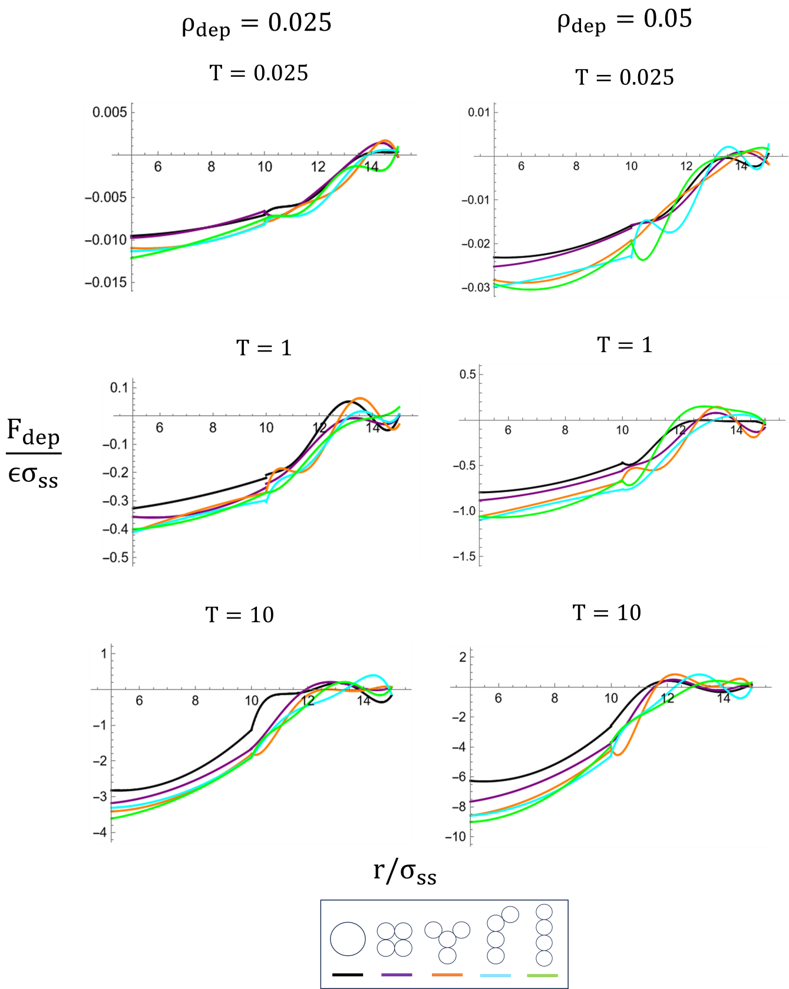}
	\caption{\label{fig:FOR1} Mean depletion forces at temperature $T=0.025$ intermediate densities ($\rho_{dep}=0.025, 0.05$) and in a wide temperature range ($T=0.025, 1, 10$) for different shapes according to the colormap.}
\end{figure}
\begin{figure}[ht] 
\includegraphics[width=0.9\linewidth]{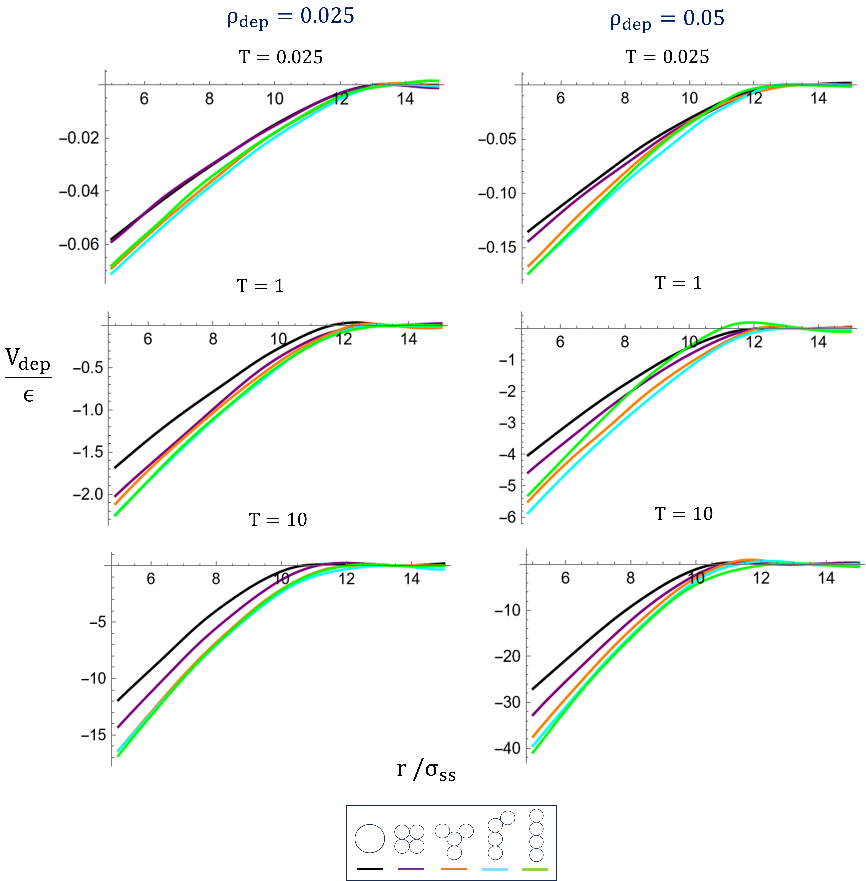}
	\caption{\label{fig:POT1} Depletion potentials at intermediate densities ($\rho_{dep}=0.025, 0.05$) and in a wide temperature range ($T=0.025, 1, 10$) for different shapes according to colormap.}
\end{figure}
\textcolor{white}{facil} The depletion forces and potentials of the different shapes are presented in Fig.\ref{fig:FOR1} and Fig.\ref{fig:POT1}, respectively, at intermediate densities and a wide temperature range. Moving from low to high temperature, the particles change their interaction from rigid to soft which affects one's ability to discern between the shape polydispersity via the depletion potential. Identical depletion profiles appear for nonidentical shapes. For example, at low temperature in intermediate density, the depletion potential is almost purely entropic since energy plays minimal role in entropy-driven effects in the hard-core limit. In this regime, disks and squares are almost identical depletion-wise. However, as temperature increases, there is an enhanced interpenetration between disks and both entropy and energy considerations matter. In this regime, depletion of squares overcomes the depletion of disks. Therefore, while at low temperature the two shapes are identical, at high temperature, they are distinct. Another example regarding elongated shapes. At low temperature: Ys, hockeys and rods are identical, while at higher temperature the depletion identity of rods separates from the Ys and hockeys. Importantly, this shows that when one intensifies the softness, one can discern better between the shapes.
\newline
\textcolor{white}{facil} The depletion forces for the highest density in this study are presented in Fig.\ref{fig:FOR2}. There the oscillating tails have high amplitudes, these forces were not integrated into potentials to avoid potentials with large errors. Thus, I decided to present the mean forces in this regime as they are. 
\section{Discussion}
I studied two-body depletion potentials induced by a fluid of small particles of different shapes at a a wide range of densities and temperatures. While the AO potential approximates well the low density computational results, it deviates as density increases. As particles modify their shape, gradually from disks to elongated, the depletion forces increase. Elongated shapes at high temperatures result in the strongest depletion effect. An oscillating tail appears at distances above big particle diameter for intermediate and high densities. Tail's amplitude is shape-dependent: minimal for compact shapes and maximal for elongated shapes. Very similar depletion potential can appear for different shapes and overcome the qualitative differences between them. An observation that depends on the depletant rigidity. This phenomenon captivates the intricate interplay of energy and entropy in soft mixtures. 
\newline
\textcolor{white}{facil} A critical point to discuss is that while I measured the two-body depletion potentials, these potentials cannot be accurately mapped onto a corresponding two-component (2C) mixture of many big particles and many small particles. Once there are more than two big particles, the many-body contributions that are absent in my work play an important role. In the case of three colloidal particles, the three-body interactions present a considerable contribution to the total interaction energy and should therefore be taken into account \cite{Brunner2004} and in principle also higher-order terms have to be considered. At high enough densities, many-body effects are not negligible, may lead to notable effects, such as a shift of the melting line. Nevertheless, the two-body potential is the expected and accepted fundamental starting point in studying the depletion interactions.  
\newline
\textcolor{white}{facil}The novelty in this work is its assessment of the interplay between shape, density and temperature in mixtures of small and big particles interacting via soft repulsive potential. An increased deviation from spherical shape results in enhanced depletion attraction. These findings can generalize to a large set of systems where entropy drives organization and phase separation, for example liquid-liquid phase separation of viruses or proteins where there is sufficient polydispersity for depletion phase separation and also explicit shape diversity. Furthermore, instead of a single shape, a mixture of different shapes at different proportions can be studied. As well, to investigate the depletants as self-propelled particles with a varying degree of activity via molecular dynamics or advanced Monte Carlo simulations. Such a study can exhibit new emerging structures as a result of an interplay between shape-dependent features to active matter features, such as motility-induced phase separation (MIPS). 
\newline
\textcolor{white}{facil} Experiments of crowded molecular systems where the change of fluid particles shape can be regulated, in a gradual manner, for example by a chemical reaction, can be performed. Then, to observe if and how the change of shape intensifies or suppresses the formation and stability of entropy-driven crystals or coarcevates. In addition, Y particles can be modelled with more details that appear in anti-bodies \cite{Nishikawa2008}. Zoo of shapes will keep pushing soft matter to ever new areas.
\section{Acknowledgments} 
{I.A. acknowledges critical and fruitful discussions with Profs Giorgio Cinacchi and Marjolein Dijkstra, and his time in The Faculty of Science and Technology in Leioa.}  \newline
\bibliography{references}
%\printbibliography
\end{document}